\input{aipcheck}

\documentclass[
    ,final            
  ]
  {aipproc}

\layoutstyle{6x9}

\begin{document}

\def\nat{Nature}%
\def\apj{ApJ}%
\def\apjl{ApJ}%
\def\apjs{ApJS}%
\def\aj{AJ}%
\def\aap{A\&A}%

\title{X-ray vs. H$_2$O maser emission in AGN}


\author{Paola Castangia}{
  address={INAF-Osservatorio Astronomico di Cagliari, Loc. Poggio dei Pini, Strada 54, 09012 Capoterra (CA), Italy}
}

\author{Avanti Tilak}{
  address={Harvard-Smithsonian Center for Astrophysics, 60 Garden Street, Cambridge, MA 02138, USA}
}

\author{Matthias Kadler}{
  address={Dr. Remeis-Sternwarte \& ECAP, Sternwartstrasse 7, 96049 Bamberg, Germany}, 
  altaddress={CRESST/USRA/NASA Goddard Space Flight Center, Greenbelt, MD 20771, USA}
}

\author{Christian Henkel}{
  address={Max-Planck-Institut f\"{u}r Radioastronomie, Auf dem H\"{u}gel 69, 53121 Bonn, Germany}
}
 
\author{Lincoln Greenhill}{
  address={Harvard-Smithsonian Center for Astrophysics, 60 Garden Street, Cambridge, MA 02138, USA}
}

\author{Jack Tueller}{
  address={NASA Goddard Space Flight Center, Greenbelt, MD 20771, USA}
}

\begin{abstract}
Correlations between X-ray and water maser emission in AGN have been recently reported. However, the lack of systematic studies affects the confidence level of these results. In the following, we introduce a project aimed at studying all the water maser sources believed to be associated with AGN activity through X-ray data obtained with the XRT and BAT instruments on-board the {\it Swift} satellite. Preliminary results of this work indicate a promising rate of XRT detections allowing us to refine follow-up observing strategies focused on investigating the nuclei of individual galaxies and deriving, on statistical basis, the main characteristics of water maser hosts. In addition, a cross-correlation between our sample and the BAT 22-months all-sky survey provides an exceptionally high detection rate at hard X-ray energies when compared to other AGN-related catalogs.
\end{abstract}

\maketitle


\section{Introduction}
The observational characteristics of active galactic nuclei (AGN), according to the leading models, are the result of the release of huge amounts of energy and angular momentum in a relatively small volume as the material accretes onto a dense central object, believed to be a supermassive black hole. The study of the detailed geometry, the kinematics and the excitation of the gas in the immediate vicinity of central engines in AGN is crucial to understand the physics of these objects. However, a comprehensive investigation of the gas in the inner parsecs of AGN is challenging because of the high angular resolution which is required to resolve structure of parsec- or subparsec-scale and the heavy obscuration of these regions at optical and ultraviolet wavelengths, due to large column densities of gas and dust along the line of sight. For these reasons, the most suitable tools for a detailed investigation of AGN are radio and X-ray observations. The 1.3 cm radio emission from H$_2$O masers originates at a few parsecs or less from the nuclear engines tracing circumnuclear accretion disks (``disk-masers'', e.\,g. NGC~4258: \cite{n4258miyo}), the inner part of relativistic jets (``jet-masers'', e.\,g. Mrk348: \cite{mrk348}) or nuclear outflows (Circinus: \cite{circinus}). Providing bright and compact hotspots, water masers can be imaged at milliarcsec resolution using Very Long Baseline Interferometry (VLBI) and are a unique instrument to map accretion disks, to estimate black hole masses, Eddington luminosities, and accretion efficiencies (see \cite{greenhill_iau07} and references therein). 
On the other hand, X-ray photons are produced in regions even closer to the supermassive central object (e.\,g. \cite{risaliti07}). X-ray spectroscopy aids estimation of absorbing column densities and, hence, of intrinsic luminosities which are a direct measure of the nuclear activity. Therefore, combining the information provided by these two tracers (H$_2$O maser and X-ray emission) has the potential to substantially improve our understanding of the physics of AGN. H$_2$O maser sources associated with active galactic nuclei tend to show a high column density or are even Compton-thick ($N_{\rm H}>10^{24}$\,cm$^{-2}$). Indeed, a recent study covering a sample of 42 masers in AGN reports that 95\% of the objects show $N_{\rm H}>10^{23}$\,cm$^{-2}$ and 60\% $N_{\rm H}>10^{24}$\,cm$^{-2}$ \cite{greenhill08}. The percentage of Compton-thick AGN rises up to 76\% when the sub-sample of disk-masers (16 objects) is considered. If confirmed, this correlation suggests that radio surveys to find water maser sources can be an alternative and less time-consuming way to identify potential obscured AGN with respect to pointed X-ray observations \cite{greenhill08}. This would be particularly important because, although synthesis models of the cosmic X-ray background predict a significant fraction of obscured AGN \cite{gilli07}, they have been, so far, very elusive objects to find \cite{dellaceca08}. A rough correlation has been also found between maser isotropic luminosity and unabsorbed X-ray luminosity \cite{kondratko_x}. Furthermore, a study on a small sample of 8 disk-masers indicates that the inner radius  and the shape of the masing disks may indeed be related to the intrinsic X-ray luminosity \cite{tilak08}.

Although promising, these studies have been affected by a lack or poor quality of X-ray data for a large percentage (~50\%) of the known maser galaxies. Hence, to overcome this limitation, we have recently performed a survey of all known H$_2$O maser sources in AGN using the {\it Swift} satellite that aims at obtaining detailed X-ray information on the largest possible number of water maser sources. This information is essential to clarify, on a firm statistical basis, the interplay between X-ray and maser emission.

\section{Sample, observations, and data reduction}
Our sample is comprised of all known H$_2$O maser sources which are believed to be associated with the nuclear activity of AGN. These include all known ``megamasers'',  namely, according to the standard definition, those sources with isotropic luminosities $L_{{\rm H_2O}} > 10$\,L$_{\odot}$ that are found in all interferometic studied cases to be associated with AGN. Furthermore, the sample also contains 9 ``kilomasers'' (sources with $L_{{\rm H_2O}} < 10$\,L$_{\odot}$ usually associated to star formation) for which the  origin of the maser emission is still uncertain (like, e.\,g., for NGC~520: \cite{castangia08}) or that show some evidence for a connection with AGN (e.\,g. M51: \cite{hagiwara07}). The total number of targets is 95.

All sources have been observed between January 2008 and November 2009 with the X-ray Telescope (XRT, 0.2--10\,keV) and the Ultra-Violet/Optical telescope (UVOT, 170--650\,nm) on-board {\it Swift}, for an average of $\sim$10\,ks for each target. Our project is mostly focused on the data taken with the XRT. The data reduction was performed using the standard pipeline package (XRTPIPELINE v. 0.12.1) to produce screened event files. Images and spectra were extracted only from data taken in Photon Counting (PC) mode, using the XSELECT v. 2.4 software. Each spectrum was extracted within a circular region of 20 pixel radius (corresponding to 47$^{\prime\prime}$), which encloses the 90\% of the Point Spread Function (PSF) at 1.5\,keV, centered at the pointing position. The background instead, was taken from a source-free circular region of radius 40 pixels. We employed the response matrices v. 011 and created the individual ancillary response files ({\it arf} files) using the task XRTMKARF v. 0.5.6. Spectral analysis was performed using XSPEC 11.3.2. In order to apply the $\chi^2$ statistics, the spectra were binned using GRPPHA to obtain a minimum of 5 counts per bin. We have also cross-correlated our maser source list with the data of the Burst Alert Telescope, (BAT, 15--150\,keV) 22-months all-sky survey \cite{tueller09}. 

\section{Preliminary results and discussion}


\begin{figure}
  \includegraphics[scale=0.6]{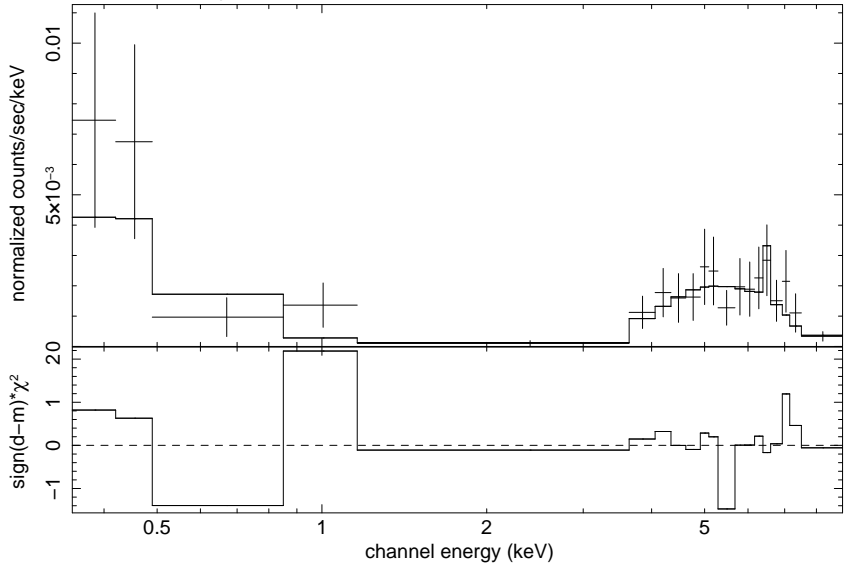}
  \caption{{\it Swift}-XRT 10\,ks spectrum of IRAS~16288+3923.}
  \label{fig:spec}
\end{figure}

Although the data reduction is still on-going, promising results have been already achieved.

We have successfully detected about 60\% (33/59) of the targets with the XRT with a signal-to-noise ratio (SNR) larger than 5 and increased the number of H$_2$O maser sources with available X-ray data by 20\%. In Fig.~\ref{fig:spec} we show a sample case of our XRT spectra for the source IRAS~16288+3929. This source is an infrared luminous Seyfert 2 galaxy and this constitutes the first detection of X-ray emission from this object. 
The spectrum indicates the presence of a soft component together with hard emission above 3\,keV. A line complex appears to be present at $E \sim$6--7\,keV. We have found that the soft emission can be well described by a black body model with temperature $kT=0.09^{+0.07}_{-0.03}$\,keV and no absorption in excess of the Galactic value, while the hard component can be modeled with an absorbed power law plus a narrow Gaussian emission line. The best fit model was obtained for a typical photon index of 1.6 (fixed) and an absorbing column density $N_\mathrm{H}=34_{-10}^{+9} \times 10^{22}$cm$^{-2}$. Although the statistics were not sufficient to constrain the line parameters, the best fit was achieved for a rest energy of 6.7\,keV, which is consistent with the K$_{\alpha}$ emission from ionized iron. 

By cross-correlating our target list with the data of the BAT 22-months all-sky survey, we have identified $\sim$30\% (26/91) of our maser sources among the objects detected by BAT at hard X-ray energies (E$>$15\,keV) with SNR$>$3. Noticeably, the majority of them (22/26) have SNR$>$4.5 and have been detected also by the blind search algorithm used to find the objects in the BAT 22-months catalog \cite{tueller09}. When compared with cross-correlations between BAT detections and any other AGN input catalog, our fraction of detections is extremely high, indicating the exceptional properties of our sample of water maser galaxies.
 

\begin{theacknowledgments}
We wish to thank C. Pagani for his valuable assistance with the XRT data reduction and F. Panessa for useful discussions on spectral fitting procedures.
\end{theacknowledgments}



\bibliographystyle{aipprocl} 

\bibliography{xray2009_proc}

\begin{thebibliography}{10}
\providecommand{\enquote}[1]{``#1''}
\expandafter\ifx\csname url\endcsname\relax
  \def\url#1{\texttt{#1}}\fi
\expandafter\ifx\csname urlprefix\endcsname\relax\def\urlprefix{URL }\fi

\bibitem{n4258miyo}
M.~{Miyoshi}, J.~{Moran}, J.~{Herrnstein}, L.~{Greenhill}, N.~{Nakai},
  P.~{Diamond}, and M.~{Inoue}, \emph{\nat} \textbf{373}, 127--+ (1995).

\bibitem{mrk348}
A.~B. {Peck}, C.~{Henkel}, J.~S. {Ulvestad}, A.~{Brunthaler}, H.~{Falcke},
  M.~{Elitzur}, K.~M. {Menten}, and J.~F. {Gallimore}, \emph{\apj}
  \textbf{590}, 149--161 (2003).

\bibitem{circinus}
L.~J. {Greenhill}, R.~S. {Booth}, S.~P. {Ellingsen}, J.~R. {Herrnstein}, D.~L.
  {Jauncey}, P.~M. {McCulloch}, J.~M. {Moran}, R.~P. {Norris}, J.~E.
  {Reynolds}, and A.~K. {Tzioumis}, \emph{\apj} \textbf{590}, 162--173 (2003).

\bibitem{greenhill_iau07}
L.~J. {Greenhill}, \enquote{{Masers in AGN environments},} in \emph{IAU
  Symposium}, edited by J.~M. {Chapman}, and W.~A. {Baan}, 2007, vol. 242 of
  \emph{IAU Symposium}, pp. 381--390.

\bibitem{risaliti07}
G.~{Risaliti}, M.~{Elvis}, G.~{Fabbiano}, A.~{Baldi}, A.~{Zezas}, and
  M.~{Salvati}, \emph{\apjl} \textbf{659}, L111--L114 (2007).

\bibitem{greenhill08}
L.~J. {Greenhill}, A.~{Tilak}, and G.~{Madejski}, \emph{\apjl} \textbf{686},
  L13--L16 (2008).

\bibitem{gilli07}
R.~{Gilli}, A.~{Comastri}, and G.~{Hasinger}, \emph{\aap} \textbf{463}, 79--96
  (2007).

\bibitem{dellaceca08}
R.~{Della Ceca}, A.~{Caccianiga}, P.~{Severgnini}, T.~{Maccacaro},
  H.~{Brunner}, F.~J. {Carrera}, F.~{Cocchia}, S.~{Mateos}, M.~J. {Page}, and
  J.~A. {Tedds}, \emph{\aap} \textbf{487}, 119--130 (2008).

\bibitem{kondratko_x}
P.~T. {Kondratko}, L.~J. {Greenhill}, and J.~M. {Moran}, \emph{\apj}
  \textbf{652}, 136--145 (2006).

\bibitem{tilak08}
A.~{Tilak}, L.~J. {Greenhill}, C.~{Done}, and G.~{Madejski}, \emph{\apj}
  \textbf{678}, 701--711 (2008).

\bibitem{castangia08}
P.~{Castangia}, A.~{Tarchi}, C.~{Henkel}, and K.~M. {Menten}, \emph{\aap}
  \textbf{479}, 111--122 (2008).

\bibitem{hagiwara07}
Y.~{Hagiwara}, \emph{\aj} \textbf{133}, 1176--1186 (2007).

\bibitem{tueller09}
J.~{Tueller}, W.~H. {Baumgartner}, C.~B. {Markwardt}, G.~K. {Skinner}, R.~F.
  {Mushotzky}, M.~{Ajello}, S.~{Barthelmy}, A.~{Beardmore}, W.~N. {Brandt},
  D.~{Burrows}, G.~{Chincarini}, S.~{Campana}, J.~{Cummings}, G.~{Cusumano},
  P.~{Evans}, E.~{Fenimore}, N.~{Gehrels}, O.~{Godet}, D.~{Grupe},
  S.~{Holland}, J.~{Kennea}, H.~A. {Krimm}, M.~{Koss}, A.~{Moretti},
  K.~{Mukai}, J.~P. {Osborne}, T.~{Okajima}, C.~{Pagani}, K.~{Page},
  D.~{Palmer}, A.~{Parsons}, D.~P. {Schneider}, T.~{Sakamoto}, R.~{Sambruna},
  G.~{Sato}, M.~{Stamatikos}, M.~{Stroh}, T.~N. {Ukwatta}, and L.~{Winter},
  \emph{ArXiv e-prints}  (2009).

\end{thebibliography}

\IfFileExists{\jobname.bbl}{}
 {\typeout{}
  \typeout{******************************************}
  \typeout{** Please run "bibtex \jobname" to optain}
  \typeout{** the bibliography and then re-run LaTeX}
  \typeout{** twice to fix the references!}
  \typeout{******************************************}
  \typeout{}
 }

\end{document}